# tRNA-isoleucine-tryptophan Composite Gene

Zhumur Ghosh [a], Jayprokas Chakrabarti [a,b,*], Bibekanand Mallick [a], Smarajit Das [a], Satyabrata Sahoo [a] and Harmeet Singh Sethi [c]

**(a) Computational Biology Group**
  Theory Department
  Indian Association for the Cultivation of Science
  Calcutta 700032, India

**(b) Gyanxet**
  BF-286, Salt Lake
  Calcutta 700064, India.

**(c) Biotechnology Department**
  Institute of Technology and Marine Engineering
  Jhinga, Diamond Harbour Road, Amira
  South 24 Parganas, WB 743 368, India.

**\* Author for correspondence**
  Telephone: +91-33-24734971, ext. 281(Off.)
  Fax: +91-33-24732805
  E-mail: tpjc@iacs.res.in ; biogyan@vsnl.net

**Abstract:**

Transfer-RNA genes in archaea often have introns intervening between exon sequences. The structural motif at the boundary between exon and intron is the bulge-helix-bulge. Computational investigations of these boundary structures in *H. marismortui* lead us to propose that tRNA-isoleucine and tRNA-tryptophan genes are co-located. Precise insilico identification of the splice-sites on the bulges at the exon-intron boundaries conduce us to infer that a single intron-containing composite tRNA-gene can give rise to more than one gene product.

**Introduction:**

In our recent work [1] we analysed 1001 cytoplasmic tRNA genes (tDNAs) in 22 archaeal genomes in search of identity elements of aminoacylation [2]. As these investigations progressed we observed some tDNAs missing in archaea [3]. In this paper we investigate *Haloarcula marismortui* ATCC 43049 (**NC 006396**) and report that a rare, and so far unnoticed, tRNA-isoleucine(UAU) gene is co-located with tDNA-tryptophan(CCA). The exon-intron boundaries are analysed [4]; the multiple splice-sites on the bulges are identified insilico.

*H. marismortui* is a halophilic archaeal isolate from the Dead Sea. The genome is organized into nine circular replicons of varying G+C compositions ranging from 54% to 62%. This halophilic archaea use the strategy of high surface negative charge of folded proteins as means to circumvent the salting-out phenomenon in a hypersaline cytoplasm [5].

Introns are found in tDNAs in all three domains of life [6]. These were first discovered in yeast's nuclear tDNA [7]; later found in several archaeal tDNAs between nucleotide-positions 37 and 38, located in anticodon (AC) loop [8]. These are the canonical introns. Archaea [9], an intermediate between Eukarya and Bacteria, have tRNAs [10] that share many similarities with either or both these domains. Archaeal tDNAs also have introns at positions other than canonical. These unusually located noncanonical introns in archaeal tDNAs were observed in 1987 [11]. Presence of intron in tDNA at canonical position is detected by the algorithm tRNAscan-SE [12]. But detecting the unusually located noncanonical introns could be difficult. This is due to the lack of prior knowledge regarding their lengths and exact locations. An incomplete set of tDNAs within the genome is generally an indication that the missing tDNAs contain one or two unusually located introns.

Eukaryotic and archaeal tRNA-gene splicing mechanism share deep similarities. Intron splicing in both these domains occur through endonucleolytic cleavage of tRNA genes producing $5'$ hydroxyl and $2', 3'$ cyclic phosphate termini at intron-exon boundaries [13]. There are at least three forms of tRNA endonuclease in archaea: homotetrameric, homodimeric, and heterotetrameric. Crenarchaeal and nanoarchaeal endonucleases are heterotetrameric. Euryarchaeal endonucleases are usually homotetrameric or homodimeric, but with exceptions. Splicing mechanism in most of the cases depends on structures that lie primarily within the introns and occurs by trans-esterification reactions. The splicing machinery cleaves introns at variable positions in pre-tRNAs within the bulge-helix-bulge, BHB, motif [14]. Splicing of introns hence is a RNA-protein interaction which requires mutual recognition of two complementary structures. Introns contribute to the specificity of splice-site recognition. Mutational analysis, secondary structure probing, and sequence analysis have shown that the conformation of the BHB motif is more important for archaeal endonuclease recognition than its sequence [15].

Mitochondrial tRNA genes were found earlier to overlap by one to six nucleotides with downstream genes on the same strand [16]. For instance, tRNA$^{Tyr}$ and tRNA$^{Cys}$ of human mitochondrial genome overlap with one another by one nucleotide at the first base of tRNA$^{Cys}$. This nucleotide is the discriminator base of tRNA$^{Tyr}$ [17, 18]. But tDNA-overlaps in *H. marismortui is* an altogether new phenomena. Here the domain of overlap is far wider, encompassing the entire tDNAs. These tDNAs are entirely co-located on the genome.

**Methods:**

We developed a computational approach to search tRNA genes having introns at positions other than canonical to identify the tRNA genes missed out by tRNAscan-SE [12] and ARAGORN [19]. About one thousand tRNA-genes from archaea were studied for this purpose. From this database of 1000 tRNA-genes we fine-tuned the strategy to locate noncanonical introns. The salient features were :(i) sequences were considered that gave rise to the regular cloverleaf secondary structure. (ii) Conserved elements: T8 (except Y8 in *M. kandleri*), G18, R19, R53, T44, Y55, and A58 were considered as conserved bases for all archaeal tRNA. Further there were tRNA-specific conserved or identity elements [1] of archaea, (iii) the promoter ahead of the $5^/$ end looked for, (iv) the constraints of lengths of stems of regular tRNA A-arm, D-arm, AC-arm and T-arm were 7, 4, 5 and 5 bp respectively. In few cases the constraints on lengths of D-arm and AC-arm were relaxed. (v) Base positions optionally occupied in D-loop were 17, 17a, 20a and 20b. (vi) The extra arm or V-arm was considered for type I and II tRNAs. The constraint on length of V-arm: less than 21 bases (vii) Noncanonical introns were considered at any position in tRNA-genes. The introns constrained to harbour the Bulge-Helix-Bulge (BHB) secondary structure for splicing out during tRNA maturation .The minimum length of intron allowed was 6 bases. (viii) The locations of the noncanonical intron and of the splice-sites on the bulges fixed to have the right conserved and identity bases for the two overlapped tDNAs. With these features in our algorithm, we extracted a rare copy of isoleucine tRNA gene missed out in *H. marismortui*. We found it embedded with tryptophan tDNA .

**Results and Discussions:**

**tRNA$^{Ile}$/ tRNA$^{Trp}$ embedded genes**: We investigated the genome of the archaea *H. marismortui*. In this archaea, the rare isoleucine tRNA gene, with anticodon UAU was not annotated earlier. We identified it between 646395 to 646575 on the genome. The entire sequence of this tRNA gene is shown in Fig 1. The intronic segments are marked in a different colour. Interestingly, when we carefully analyse and observe the specific patterns of this genomic segment, it has features of both tRNA-isoleucine (UAU) and tRNA-tryptophan (CCA) genes. Our observations lead us to propose two alternate intron splice sites giving rise two different gene products from this single composite tRNA gene. We present the two optimized secondary structures arising from this gene, after the introns are spliced out, in Fig 2a and 2b. One of these correspond to tRNA-isoleucine; the other one tRNA-tryptophan. We now analyse insilico the secondary structure at the exon-intron boundary – the bulge-helix-bulge (BHB). To be precise a relaxed BHB, having reverse complementary features, is found to be more appropriate. It is the conformational structure most easily recognized and processed by archaeal splicing mechanism. We elucidate this identification now. Note that intron splicing occurs during tRNA maturation following the transcription of the gene. Splicing in archaea is enzyme catalyzed, initiated by an endonuclease that excises the intron to yield half-molecules with ends containing a 2$^/$ -3$^/$ cyclic phosphate and a 5$^/$ - OH.

The secondary structure of the intronic sequence with a periodic repeat of the motif "helix (h)-bulge (b)-helix (h)" is shown in Fig 3. The first one designated as $h_1$ is a 5 base pair helix; $h_2$ is a 4 base pair helix. $h_3$, $h_4$ and $h_5$ are 9bp, 10bp and 8bp helices respectively with one nucleotide opening in each of these 3 helices. The first bulge $b_1$ is a 3 nucleotide (nt) bulge. $b_2$, $b_3$, $b_4$ and $b_5$ are 14, 4, 20 and 7 nt bulges respectively. Among these bulges, splice sites for the noncanonical intron (NCI) of isoleucine tRNA lie on the opposite of $b_1$ and on $b_4$ between adenine-guanine and uracil-guanine respectively. The splice sites for canonical intron (CI) of the same lies on the same two bulges between adenine-cytosine and uracil-adenine. To obtain isoleucine-tRNA as one of the gene products noncanonical splicing are proposed at the positions marked "IleN$_s$" and "IleN$_e$" on Fig 3; the canonical splicing proposed at "IleC$_s$" and "IleC$_e$", keeping the rest of the gene intact. For the other gene product, tryptophan-tRNA, we propose only a canonical splicing at "TrpC$_s$" and "TrpC$_e$". The end points of canonical splice positions for both tryptophan and isoleucine in this composite tRNAs gene coincide. Following the standard tRNA numbering nomenclature we have the following: for isoleucine tRNA, the noncanonical intron lies between $32^{nd}$ and $33^{rd}$ tRNA-positions. It has length 15. Then there is the canonical intron between $37^{th}$ and $38^{th}$ tRNA-nucleotide positions. The canonical intron is 21 bases long. For the other gene product, namely tRNA-tryptophan, there is just the canonical intron between 37th and $38^{th}$ tRNA-positions. This canonical intron is of length 36.

The tRNA-isoleucine has the important conserved bases, adenine at $73^{rd}$ and $35^{th}$ positions and Uracil at $36^{th}$, necessary for aminoacylation by isoleucine aminoacyl tRNA synthetase. It has all the conserved bases and base-pairs of other archaeal isoleucine tRNAs. The tryptophan-tRNA product has all the important features of tryptophan tRNA. Cytosine at $34^{th}$ and $35^{th}$ positions and adenine at $36^{th}$ position are the identity elements for proper aminoacylation, in addition to the discriminator base, adenine, at 73.

**Conclusions:**

In archaea we studied the availability of all tRNAs over the entire set of sequenced genomes. Extending our investigation to the newly sequenced halophile *H. marismortui* we identified the non-annotated isoleucine- tRNA gene in it. Interestingly, this tRNA gene is co-located with tRNA-tryptophan. In some of the primary transcripts of mitochondrial tRNA of animals, tRNA genes are known to overlap by one to several bases. But here the overlap encompasses the entire transfer-RNA genes. This is a novel phenomena . Splicing of introns at alternate positions on the bulges of the BHB at the intron-exon boundary generates tRNA-Ile (UAU) or tRNA-Trp(CCA). This insilico evidence leads us to propose that *a single intron-containing composite tRNA-gene can give rise to two tRNA products.* Assuming this alternate intron splicing mechanism we speculate now on how it works. There are at least two possibilities. One is to assume that tRNA- endonuclease recognition sites are not uniquely determined. Coded into the tDNA there are sequence/structural signatures that lead to one or the other of the splice-sites. An alternate possibility is to assume that there exist two competing almost equally stable structural motifs   leading to two alternate splicing modes.

**References**:

[1] B. Mallick, J.Chakrabarti, S. Sahoo, Z. Ghosh and S. Das, Identity Elements of Archaeal tRNA, DNA Research (article in press) (2005).

[2] S. Fukai, O. Nureki, S. Sekine, A. Shimada, D.G. Vassylyev and S. Yokoyama, Mechanism of molecular interactions for tRNA$^{Val}$ recognition by valyl-tRNA synthetase, RNA 9 (2003) 100-111.

[3] B. Mallick, J. Chakrabarti, S. Das and Z. Ghosh, tRNomics: A comparative analysis of *Picrophilus torridus* with other archaeal thermoacidophiles, IJBB 42 (2005) 238-242.

[4] T.Yoshihisa, K.Yunoki-Esaki, C. Ohshima, N. Tanaka, and T. Endo, Possibility of cytoplasmic pre-tRNA splicing: the Yeast tRNA splicing endonuclease mainly localizes on the mitochondria, MBC 14 (2003) 3266-3279.

[5] N.S. Baliga, R. Bonneau, M. T. Facciotti, M. Pan, G. Glusman, E.W. Deutsch, P. Shannon, Y. Chiu, R.S. Weng, R.R. Gan, P. Hung, S.V. Date, E. Marcotte, L. Hood and W.V. Ng, Genomic sequence of *H. marismortui* : A halophilic archaeon from the Dead sea, Genome Research 14 (2004) 2221-2234.

[6] J.L. Diener and P.B. Moore, Solution structure of a substrate for the Archaeal pre-tRNA splicing endonucleases: The Bulge-Helix-Bulge motif, Molecular Cell 1 (1998) 883-894.

[7] P. Valenzuela, A. Venegas, F. Weinberg, R. Bishop and W.J. Rutter, Structure of yeast phenylalanine-tRNA genes: An intervening DNA segment within the region coding for the tRNA, Proc Natl Acad Sci USA 75 (1978) 190-194.

[8] C.J. Daniels, R. Gupta and W.F. Doolittle, Transcription and excision of a large intron in the tRNA Trp gene of an archaebacterium, *Halobacterium volcanii* , J Biol Chem 260 (1985) 3132-3134.

[9] S. Chattopadhyay, S. Sahoo, W.A. Kanner and J. Chakrabarti, Pressures in archaeal protein coding genes: A comparative study, Comparative and Functional Genomics 4 (2003) 56-65.

[10] S. Nakamura, M. Ikeguchi and K. Shimizu, Dynamical analysis of tRNA$^{Gln}$-GlnRS complex using normal mode calculation, Chem. Phys. Lett. 372 (2003) 423-431.

[11] G. Wich, W. Leinfelder, and A. Böck, Genes for stable RNA in the thermophile *Thermoproteus tenax*: introns and transcription signal, EMBO J 6 (1987) 523-528.

[12] T.M. Lowe and S.R. Eddy, tRNAscan-SE: A program for improved detection of transfer RNA genes in genomic sequence, Nucleic Acids Res 25 (1997) 955-964.

[13] J. Kjems and R. Garrett, Novel splicing mechanism for the ribosomal RNA intron in the archaebacterium *Desulfurococcus mobilis*, Cell 54(1988) 693-703.

[14] H. Li, C.R. Trotta and J. Abelson, Crystal structure and evolution of a transfer RNA splicing enzyme, Science 280(1998) 279-284.

[15] J. Lykke-Andersen and R.A. Garrett, Structural characteristics of the stable RNA introns of archaeal hyperthermophiles and their splicing junctions, J Mol Biol 243 (1994) 846-855.

[16] S .–i. Yokobori and S. Pääbo, Transfer RNA editing in land snail mitochondria , Proc Natl Acad Sci USA 92(1995) 10432-10435.

[17] A. Reichert, U. Rothbauer and M. Mörl, Processing and editing of overlapping tRNAs in human mitochondria, J Biol Chem 278(1998) 31977-31984.

[18] S. Binder, A. Marchfelder and A. Brennicke, RNA editing of tRNA[Phe] and tRNA[Cys] in mitochondria of *Oenothera berteriana* is initiated in precursor molecules, Molecular Genetics and Genomics 244 (1994) 67-74.

[19] D. Laslett and B. Canback, ARAGORN, a program to detect tRNA genes and tmRNA genes in nucleotide sequences, Nucleic Acids Res 32(2004) 11-16.

Isoleucine tRNA(TAT) gene : 646395-646575
NCI between 31 & 32 and a CI

GGGGTCGTGGCCTAGTCCGGGAAGGCGGCTGACTCCAGAGGCCACGCG
CCTGGGACGACACTCCAAGGGCTGATATACTGAGCGGCCGGCTGATCAC
CGGTTCGCGACGATGACCCTCTGGAGTTCCGAGGCGCAGGACGGAGATA
TCAGCCGATCGGGGGTTCAAATCCCTCCGACCCCA

Tryptophan tRNA(CCA) gene : 646395-646575

GGGGTCGTGGCCTAGTCCGGGAAGGCGGCTGACTCCAGAGGCCACGCG
CCTGGGACGACACTCCAAGGGCTGATATACTGAGCGGCCGGCTGATCAC
CGGTTCGCGACGATGACCCTCTGGAGTTCCGAGGCGCAGGACGGAGATA
TCAGCCGATCGGGGGTTCAAATCCCTCCGACCCCA

**Figure 1**

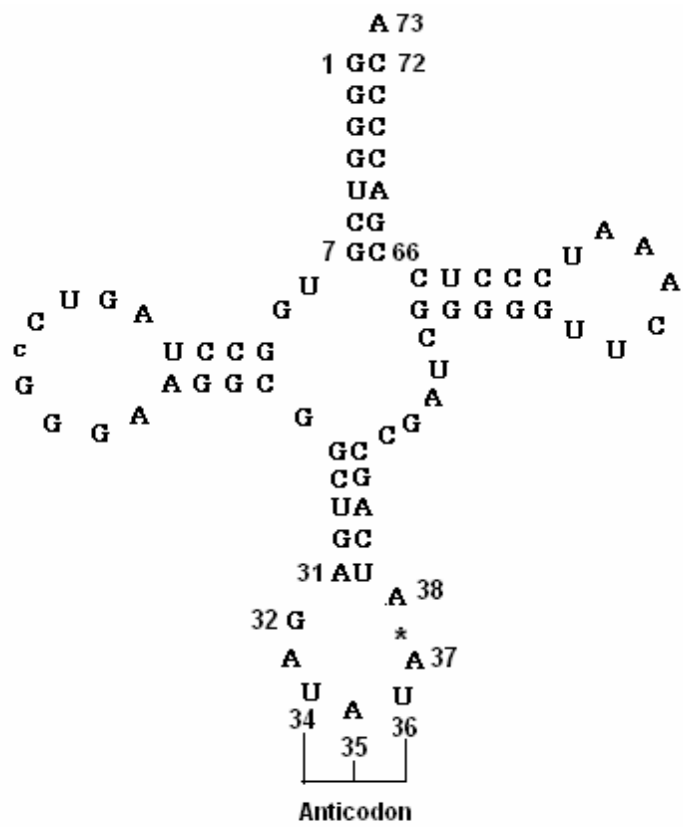

**Figure 2a**

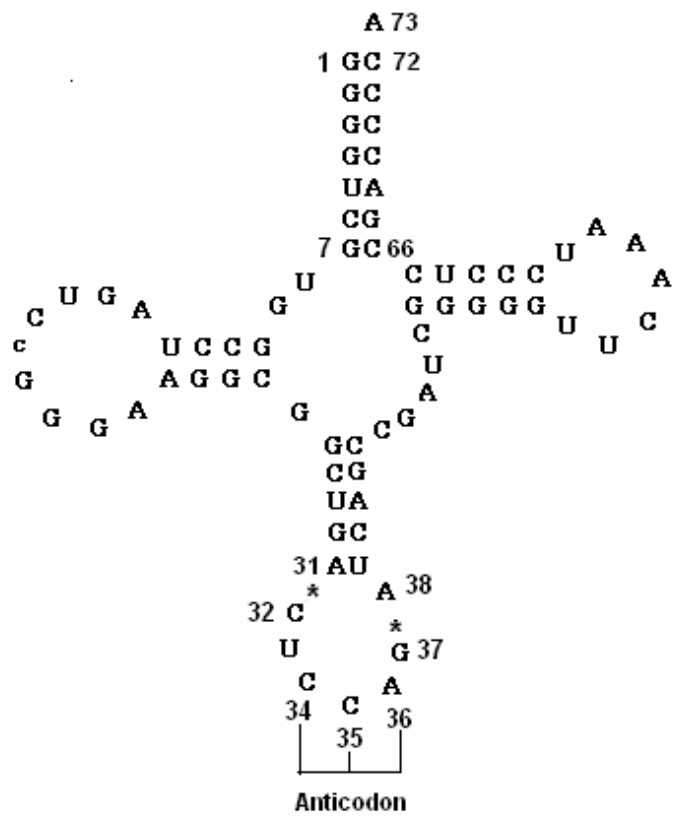

**Figure 2 b**

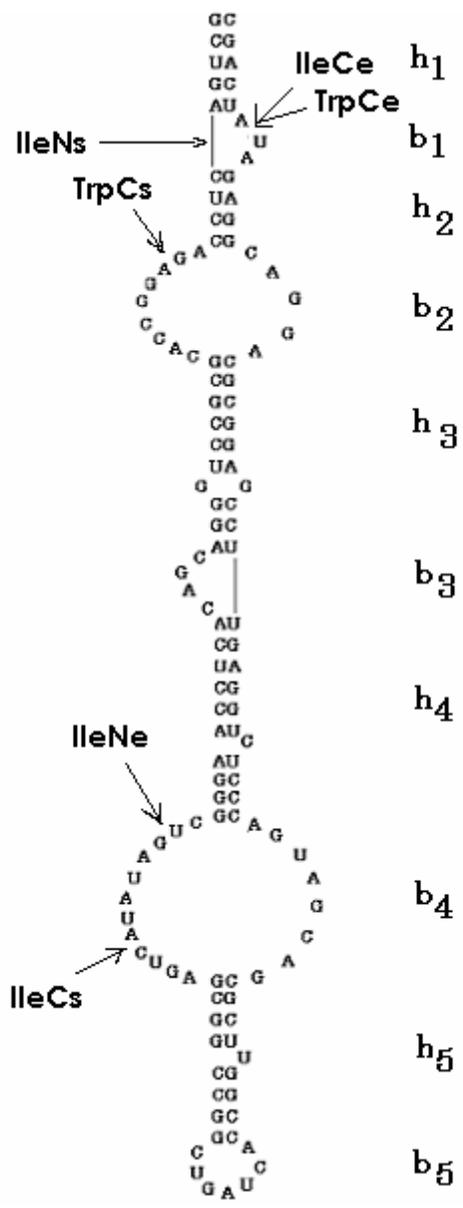

**Figure 3**

**Figure Legends**

**Figure 1**: Overlapping tRNA gene sequences of the rare tRNA$^{Ile}$(UAU) with tRNA$^{Trp}$(CCA) of *Haloarcula marismortui* . The portion in black portion denotes the gene ; blue indicates NCI ; brown indicates CI ; green indicates 34$^{th}$ , 35$^{th}$ & 36$^{th}$ nucleotides.

**Figure 2**a: Secondary structure of tRNA$^{Ile}$(UAU)

**Figure 2**b: Secondary structure of tRNA$^{Trp}$(CCA)

**Figure 3**: BHB structure of overlapped tRNA$^{Ile}$(UAU) and tRNA$^{Trp}$(CCA)

Ns: Noncanonical intron start position; Ne: Noncanonical intron end position; Cs: Canonical intron start position; Ce: Canonical intron end position.

⟶        This signifies splicing sites on the introns in tDNAs h denotes helix and b denotes bulge and these are assigned numbers like 1, 2, 3 etc.